\documentclass[11pt,a4paper]{article}
\pdfoutput=1

\usepackage{jheppub}
\usepackage{slashed}

\usepackage{amsmath}
\usepackage{amssymb}
\usepackage{amsfonts}

\newcommand{\dprime}{{\prime\prime}}
\newcommand{\ud}{\textrm{d}}

\newcommand{\commut}[2]{[\,#1\,,\,#2\,]}
\newcommand{\acommut}[2]{\{\,#1\,,\,#2\,\}}

\renewcommand{\bar}{\overline}
\newcommand{\tr}{\textrm{tr}}
\newcommand{\Tr}{\textrm{Tr}}

\newcommand{\vev}[1]{\left<\,#1\,\right>}

\newcommand{\rst}[1]{\raise+.6ex\hbox{#1}}

\newtheorem{thm}{Theorem}
\newtheorem{lem}{Lemma}

\title{On two dimensional non-abelian chiral lattice gauge theories in Ginsparg-Wilson formalism}

\author{Yanwen Shang}
\affiliation{Perimeter Institute for Theoretical Physics, \\
31 Caroline St. N., Waterloo, ON, Canada, N2L 2Y5}

\emailAdd{yshang@perimeterinstitute.ca}

\abstract{Defining chiral lattice gauge theories in the Ginsparg-Wilson formalism
is complicated by the so-called fermion measure problem. It has been proven
for the abelian theories that smooth well-behaved fermion measure exists if
and only if the anomaly-free condition is granted, and the same was shown
to hold in perturbative theories for non-abelian gauge groups, but the
non-perturbative proof is absent. In this paper, we consider 
a simpler problem in $2$-d and present a proof 
for the existence of smooth and gauge invariant fermion measure on the
gauge field configuration space with zero field strengths for arbitrary
compact Lie groups, provided the anomaly-free conditions are satisfied.
It is conjectured that such consideration is sufficient for the unknown
full proof.}

\arxivnumber{}

\begin{document}

\maketitle

\section{Introduction}
\label{sec:intro}
Defining gauge theories with chiral fermion content on a finite lattice has been
a longstanding difficult subject. The initial challenge stemmed from the
infamous "fermion-doubling problem" which leads to the multiplication
of fermion spectra in the continuum limit if simple-minded 
discretization for the Dirac operator is used.  With
the extra modes in the spectrum, they always form vector multiplets, 
preventing a lattice regularization for 
theories with chiral fermion content \cite{fujikawa}.

Various methods of removing the fermion doublers are known. Each introduces new difficulties when
solving the old one. As a general principle, explicit breaking of chiral
symmetry on a finite lattice is a necessity so that the would-be ``doublers'' 
are endowed with a mass of the order of the inverse lattice spacing and eliminated
in the continuum. Depending on the methods, it may require the fermion content 
to be vector-like to start with, certainly not a welcomed restriction for 
defining chiral theories.  In fact, the lack of
an exact chiral symmetry on a finite system obscures the 
proper definition for ``chiral theories'' after all. 

The Ginsparg-Wilson formalism \cite{Ginsparg:1981bj} stands out in this regard 
and earned itself lots of attentions from the community. 
A convenient feature of this formalism is that despite the ordinary chiral symmetry being
broken, it allows one to define a new ``chiral symmetry'' on a 
finite lattice which approaches the usual one in the continuum limit. 
With respect to this new ``chiral symmetry'', 
the so-called Ginsparg-Wilson, or the overlap, ``chiral fermions'' 
can be defined and interesting theories
for them are easily constructed, provided that the chiral symmetries
are not gauged.

Gauging the theory with Ginsparg-Wilson ``chiral fermions'' 
poses some serious new challenges.
The difficulty is often referred to as the ``fermion measure
problem''. Even without gauging, the partition function
for a chiral theory is well-defined only up to a pure phase.
As long as such an ambiguous phase is independent from all physical fields, it never
appears in the normalized correlation functions and therefore bears no
physical significance. The moment 
gauge fields are present, coupled to the Ginsparg-Wilson chiral fermions, 
as explained below, the said ambiguous phase
necessarily becomes a non-trivial functional
of the gauge field configuration, leading to serious concerns.
One must hope to find a way of defining this phase
such that it is a smooth, local, and gauge invariant functional
of the gauge fields throughout the entire space of
the so-called ``permissible gauge field configurations''. Such a choice
is referred to a ``good fermion measure'', and when it exists,
the phase ambiguity can be absorbed by adjusting the local counter-terms as 
the continuum limit is approached, a step needed in any case.
However, if such a choice fails to exist, the theory is ill defined
and the functional integral for the gauge fields does not make any sense on 
finite lattices.

It has been proven for abelian gauge theories that a good ``fermion measure'' 
exists if and only if the gauge anomaly cancellation condition (in two dimensions):
\begin{equation}\label{eq:abelian_anomaly_cancellation} 
\sum_i q_{L, i}^2=\sum_j q^2_{R, j}
\end{equation}
is satisfied \cite{Luscher:1998pqa, Luscher:1998kn, Luscher:1998du}.
Here $q_{L/R, i}$ are the charges of each
fermion flavor indexed by $i$, and $L/R$ refers to its chirality.
This intriguing result, even though not at all surprising, certainly 
shed light on yet another interesting character of the Ginsparg-Wilson 
formalism, making it a theoretically appealing subject for further 
investigations.  For the non-abelian gauge theories, however, the similar
theorem is yet to be found. A perturbative proof was given in \cite{Luscher:2000zd},
showing that to all order of the perturbative expansion, it is indeed true
that the existence of the ``good fermion measure'' coincides with the
absence of gauge anomalies, but a full non-perturbative proof remains unknown.

While a complete understanding to the aforementioned result
requires full knowledge of the permissible gauge field configuration space, a curious fact 
is that, in the abelian case, the sought coincidence can be understood to a great 
extent when most part of the gauge field configuration space
is ignored \cite{Neuberger:1998xn, Poppitz:2007tu}.
The anomaly cancellation condition emerges already 
if one studies the zero field strength configurations only.
Furthermore, focusing on the homogeneous gauge field configurations appears to be sufficient. 
Finally, if one is willing to take one, perhaps a very big
one, step backward and consider the same problem on $2$-d lattices,
the quoted theorem can be deduced with minimal efforts using
some simple geometrical considerations. Of course, we know why
this sequence of simplifications arises, 
the \emph{permissible field configuration space}
for the abelian gauge fields was found to be given by
\begin{equation}\label{eq:abelian_u}
\mathbb U[U(1)]=\mathbb U_0[U(1)]\times\mathbb F\,,
\end{equation}
where $U_0[U(1)]$ is the space consisting  of all zero field strength
configurations and the factor $\mathbb F$ is contractible.  
The space $\mathbb U_0[U(1)]$ is further given by
\begin{equation}\label{eq:abelian-u0}
\mathbb U_0[U(1)]=T^2\times U(1)^{N^2-1}\,.
\end{equation}
Here $T^2$ is a $2$-dimensional torus describing homogeneous 
field configurations on a periodic lattice,
and the remaining $U(1)$ factors correspond to gauge
transformations.  As we explain in the following sections, a line 
of reasoning leads to the conclusion that 
a ``good fermion measure'' exists on $\mathbb U$ if and only if
it does so on the $T^2$ factor.

Now, should we attempt to study the non-abelian gauge theories,
the first thing to notice is that, in $2$-d, 
the gauge anomaly cancellation condition takes a fairly
similar form as \eqref{eq:abelian_anomaly_cancellation}
\cite{hwang}, which reads
\begin{equation}\label{eq:non-abelian_anomaly_cancellation}
\sum_i \tr{\;t^a_{L, i}t^b_{L, i}}=
\sum_j \tr{\;t^a_{R, j}t^b_{R, j}}
\end{equation}
where $t^a_{L/R, i}$ are the generators of the Lie algebra $\mathfrak g$ 
for the gauge group $G$ in the representation of each fermion flavor.
A few steps of algebra show that the equality is secured
as long as it holds true within any one of the Cartan subalgebra 
$\mathfrak c\subset \mathfrak g$. So, essentially, only the
abelian subgroups in $G$ contribute. This 
observation suggests that whatever that is known for the
$2$-d abelian chiral lattice theories might be easily generalized
into the non-abelian ones. 

We take a small step toward this direction in this paper, assuming
the gauge group $G$ is compact. Given our
experience in the abelian case, we hope that studying the gauge field
configuration space corresponding to the zero field strength is sufficient.
To fully justify this simplification requires substantially more work
and we must leave it to the future. However, we can demonstrate that
the anomaly cancellation equation \eqref{eq:non-abelian_anomaly_cancellation}
does emerge already when attempts to construct a smooth and gauge invariant
fermion measure over the space $\mathbb U_0[G]$ are made. Furthermore,
just as in the abelian case, it is sufficient to construct
the measure on the subspace of $\mathbb U_0[G]$ that corresponds to
homogeneous gauge configurations only. 

The result mainly relies on the fact that for an
arbitrary compact semi-simple Lie group $G$, the space $\mathbb U_0[G]$ is
given by
\begin{equation}\label{eq:non-abelian_u0}
U_0[G]=S(G)\times G^{N^2-1}
\end{equation}
where $S(G)$ is the space of the commuting pairs $(g_1, g_2)$, $g_1, g_2\in G$,
and the $G^{N^2-1}$ factor corresponds to the gauge transformations.
The space $S(G)$ can be further expressed as the product of a pair
of the same maximal tori of $G$, denoted as $T_k^2$, foliated by gauge orbits.
Each gauge orbit appears as the conjugacy class of $G$ quotient
the Weyl group.  $T_k$ is a $k$-dimensional torus
and $k=\textrm{rank} G$.  The claim is that
finding a good fermion measure on $T_k^2$ ensures
the existence of the same on $\mathbb U_0[G]$.
To prove this, a small interesting Lemma \ref{lem:non-abelian_stokes} referred to
as the ``non-abelian Stoke's theorem'' by us, has been used.

We should mention that it might sound ridiculous that one feels
comfortable to concentrate his attention to zero field strength
gauge configurations only, since the true dynamics are all about
non-zero field strengths. This, however, is a slight misconception.
Recall that one is ultimately interested in the continuum limit
where each lattice plaquette effectively becomes a point. One certainly
would expect the Wilson-loop around a single point to be trivial, or
the gauge field is singular. Consequently, the lattice
simulation is done only over the gauge field configurations whose
Wilson-loop around each plaquette is bounded by a small number
$\epsilon$. In other words, on a finite lattice,
the permissible gauge field configuration space is 
a small bounded region surrounding the slice 
$\mathbb U_0[G]$ as we will elaborate slightly further in 
Sec. \ref{sec:anomaly_free}. While this suggests considering
the trivial-sounding space $\mathbb U_0[G]$ is not nonsensical,
it is not a rigorous proof either since $\epsilon$ is not
infinitesimal on a given finite lattice. One may hope to use
the topological nature of the proof, as explained below,
to extract a complete proof by taking the limit of
the lattice spacing and $\epsilon$ to zero continuously.

Some may also wonder if this discussion is worthwhile at all
since the perturbative proof for the non-abelian theories is known. Wouldn't
zero field strength configurations correspond to just the zero-th
order term in a perturbative expansion? The reason that
this is not true is that even when the field strengths vanish, there
are nontrivial gauge field configurations corresponding to ``large'' Wilson-loops.
In fact, from our experiences from the abelian theories, it is precisely
the considerations of these large Wilson-loops rather 
than the nonzero field strengths that lead to the emergence of
the anomaly cancellation condition.

The rest of the paper is organized as the following. We start
in Sec. \ref{sec:ginsparg-wilson}
with a brief introduction to the Ginsparg-Wilson formalism and
the so-called fermion measure problem when the theory is gauged.
In section \ref{sec:abelian} we review the solution to this
problem in the abelian case and the emergence of the anomaly-free
condition. Without complete justification, we attempt to
generalize the result to non-abelian theories by restricting the
gauge group $G$ onto one of its maximal tori in \ref{sec:non-abelian}. 
In section \ref{sec:main}, we prove the claim \eqref{eq:non-abelian_u0},
and, in Sec. \ref{sec:su2}, discuss an explicit toy example for $G=SU(2)$. 
We finish the paper with additional discussion in section \ref{sec:conclusion}.
A simple proof for a funny theorem called the ``non-abelian Stoke's
theorem'' is presented in App. \ref{app:stokes}.

\vspace{1 cm}

Let us settle the notations and terminologies in this paper.
We study chiral theories on a $2$-d square lattice
denoted as $\mathbb L$ throughout the rest of the discussions. The size
of $\mathbb L$ is always assumed to be $N\times N$. Fermions
are Grassmann fields living on the vertices whose coordinates
are specified by a pair of integers as $m=(x, y)$ and
$x, y\in [0, N-1]$. The notation $\hat\mu$ denotes the
unit vector in the $\mu$-th direction, and so $m+\hat\mu$ is the 
neighboring vertex of $m$ one unit of lattice spacing to 
its right if $\mu=1$ or above if $\mu=2$.
Links in $\mathbb L$ can be labeled by $l_\mu(x)$ 
which is the link between the vertices $x$ and $x+\hat\mu$, pointing
from the former to the latter. Gauge fields live on the links.
Depending on the representations, assumed to be unitary throughout, 
for each fermion, the corresponding
link field on $l_\mu(x)$ will be denoted as $U^{R_i}_\mu(x)$,
a unitary matrix representing the elements of $G$
in the representation $R_i$. For brevity, we often omit the 
superscript specifying the representation, particularly when attention
is paid only to a single fermion flavor, and freely refer to 
$U_\mu(x)$ either as the group element or the corresponding matrix 
representation.  One must be cautious though that on an identical
gauge field background, the corresponding matrices
for the link field differ for different fermion flavors, a detail
assumed understood implicitly in most of this paper. Gauge transformations
are generated by a group element valued function $\omega(x)$ living on the vertices, 
and the link fields transform as
\[
U\mu(x)\rightarrow \omega(x+\hat\mu)U_\mu(x)\omega(x)^{-1}\,.
\]
Occasionally, we may also refer to the links pointing in the opposite directions
as $l_{-\mu}(x)$ and correspondingly $U_{-\mu}(x+\hat\mu)=U_\mu(x)^\dag$.
We say the gauge field configuration has zero field strength if the
Wilson-loop around every plaquette in $\mathbb L$ is trivial.
Periodic boundary conditions are assumed to all fields, and
the lattice spacing is fixed to be $1$.

\section{The Ginsparg-Wilson formalism and the fermion measure problem}
\label{sec:homo}
We briefly review the Ginsparg-Wilson formalism, focusing only on
the most directly relevant results and refer the interested readers 
to the literatures \cite{Ginsparg:1981bj, Luscher:2000hn, Poppitz:2009gt}
and the references therein for more details.
We first explain the modified chiral symmetry that is exact
on finite lattices and the definition for Ginsparg-Wilson chiral 
fermions, without introducing the gauge fields,
and then present the topological question one must address for defining
consistent chiral gauge theories in the same framework,
dubbed as the ``fermion measure problem''.

\subsection{A brief review of Ginsparg-Wilson chiral fermions}
\label{sec:ginsparg-wilson}
To eliminate the doublers, the ``old chiral'' symmetry is broken
on lattices by substituting in the Lagrangian the so-called 
Ginsparg-Wilson operator in place of the Dirac operator. It obeys the
Ginsparg-Wilson relations: \footnote{While this paper is mostly
about $2$-D theories, we still label the parity flipping
matrix $\gamma_5$ for conventional reasons.}
\begin{equation}\label{eq:ginsparg-wilson}
\acommut{D}{\gamma_5}=D\gamma_5 D\,,\quad (D\gamma_5)^\dag=D\gamma_5\,.
\end{equation}
Here, $\acommut{\ast}{\ast}$ is the anti-commutator.
The first equation implies that in the continuum limit $D$ anti-commutes with
$\gamma_5$  as the right-hand side is of higher order
of momenta and vanishes in the continuum limit. The second 
ensures $D$ is Hermitian in the same limit. Now that
$D$ fails to anti-commute with $\gamma_5$, the ordinary definition
for chirality ceases to be useful. However, one may define
a ``new $\gamma_5$ operator'' by
\[
\hat\gamma_5=(1-D)\gamma_5
\]
which approaches $\gamma_5$ in the continuum for the same reason
as just mentioned, and, by the Ginsparg-Wilson relations,
\[
\hat\gamma^\dag_5=\hat\gamma_5\,\qquad\hat\gamma_5^2=1\,.
\]
So $\hat\gamma_5$ is indeed similar to $\gamma_5$, whose only
eigenvalues are $\pm 1$. Consequently $\Tr\hat\gamma_5$ is
an integer that can not vary smoothly with respect to
any continuous parameters and must vanish in the trivial 
topological sector. Thus
\[
\Tr\hat\gamma_5=\Tr'\gamma_5\,,
\] 
where $\Tr'$ refers to the regularized trace in the continuum
limit, as long as the regularization and renormalization procedure is continuous.
These equations combined led to
\begin{equation}\label{eq:hatgamma_5_D}
\hat\gamma_5 D+D\gamma_5=0\,,
\end{equation}
from which one discovers the new exact ``chiral symmetry''.
Consider, for example, the theory 
of a single Dirac fermion described by the Lagrangian $\mathcal L=\bar\psi D\psi$.
The action is manifestly invariant under the ``axial'' rotation:
\begin{equation}\label{eq:chiral}
\bar\psi\rightarrow \bar\psi e^{i\theta\hat\gamma_5}\,,\qquad
\psi\rightarrow e^{i\theta\gamma_5}\psi\,.
\end{equation}
The transformation, however, is not unitary if $\Tr\hat\gamma_5\ne 0$,
in which case, the Jacobian is given by
\[
J^{-1}=1+\theta\Tr\hat\gamma_5\,.
\]
This is the manifestation of the axial anomaly on a finite lattice.
Recall that $\Tr\hat\gamma_5=n_+-n_-$, where $n_+$ and 
$n_-$ are the number of eigen-modes for 
$\hat\gamma_5$ corresponding to the eigenvalues $\pm 1$ respectively, and,
in the continuum limit, is exactly the regularized trace of
$\gamma_5$, or the index of the Dirac operator.

Given the exact chiral symmetry just defined, theories for chiral fermions 
can be constructed using the ``chiral projection operators'':
\[
\hat P_\pm=\frac{1\pm\hat\gamma_5}{2}\,,\qquad
P_\pm=\frac{1\mp\gamma_5}{2}\,,
\]
where $\pm$ denotes the left or right-handed chiralities respectively. 
The Lagrangian for a single chiral fermion may be expressed formally
as $\mathcal L=\bar\psi \hat P_+ D P_+\psi$. 
By equation \eqref{eq:hatgamma_5_D}, one may omit either one
of the two projection operators in the Lagrangian. To define a non-vanishing
partition function, one must restrict the functional integral for $\bar\psi$
and $\psi$ to be within the $+1$-eigenspace of $\hat P_+$ and $P_+$
respectively.  More explicitly, one chooses a set of orthonormal eigenvectors 
$u_i$ and $v_i$ such that $\hat P_+ u_i= u_i$ and $P_+ v_i= v_i$,
where $i$ runs from $1$ to the half of the dimension of the Hilbert space of
both $\bar\psi$ and $\psi$, and define the partition function by
\begin{equation}\label{eq:chiral_partition}
Z=\int\prod_{i, j}\ud \bar c_i \ud c_j e^{\bar c_i c_j u_i^\dag D v_j}\,.
\end{equation}
Here, $\bar c_i$ and $c_j$ are two sets of Grassmann variables. 

Such a partition function is not uniquely defined since there exist infinitely
many choices for the orthonormal basis and equation \eqref{eq:chiral_partition}
is not independent from such freedom.
Should we choose $u'_i=\mathcal U_{ij} u_j$, where $\mathcal U_{ij}$ is
a unitary matrix, defining $Z$ as given above leads to an result
that differs by a factor of $\det(\mathcal U_{ij})$, which is a pure phase.

Without introducing the gauge fields, this phase ambiguity is easily accommodated.
In fact, it is always present in any chiral theory whenever the fermion
representation is complex. It always disappears, on the other hand,
in all physical observables determined by normalized correlation functions.

\subsection{The topological obstruction for defining chiral gauge theories}
\label{sec:ginsparg-Wilson_gauged}
Gauging the theory, however, faces some serious challenges.
With gauge fields $U_\mu(x)$ included, the Ginsparg-Wilson operator
$D$ is covariantized, of which an explicit realization is given later, 
such that under the gauge transformation
$U_\mu(x)\rightarrow \omega(x+\hat\mu)U_\mu(x) \omega(x)^{-1}$, the operator
transforms as $D_{mn}\rightarrow\omega(m) D_{mn}\omega(n)^{-1}$.
The same property is automatically shared by both $\hat P_{\pm}$
and $\hat\gamma_5$.  Once covariantized, the kinetic term for a single 
chiral fermion in a particular unitary representation of $G$ may be 
defined by
\[
\mathcal L=\bar\psi_+ D\psi_+\,,
\]
where $\bar\psi_+$ and $\psi_+$ are Grassmann valued
eigenvectors of $\hat\gamma_5$ and $\gamma_5$ with eigenvalue $+1$
and $-1$ respectively.  

Once again, to define the partition function,
an orthonormal basis $(u_i, v_i)$ is chosen, and the functional
integral \eqref{eq:chiral_partition} is uniquely determined up to a phase angle.  
In the current scenario, however, this ambiguous phase of $Z$ is
necessarily gauge field dependent simply because
the operator $\hat P_+$ is, and, as the gauge field varies, its
eigenvectors can not stay fixed and the corresponding eigenspace
rotates in a non-trivially way.

Instead of being a simple constant phase that has no physical
consequences, the phase of $Z$ is a functional of the gauge fields 
now. More generally speaking, if $\mathbb U[G]$ is the
permissible gauge field configuration space, the phase ambiguity 
of $Z$ leads to a $U(1)$ bundle over $\mathbb U[G]$ that locally 
appears as $U(1)\times V$, where $V$ is a small patch of
$\mathbb U[G]$.  Making a specific choice for the phase for
$Z$ at each gauge field configuration point amounts to finding
a global section in the said bundle, and is often referred to 
as choosing ``a fermion measure'' in the literature.

Working with the eigenvectors $u_i$ is often cumbersome and overly
complicates the problem. More convenient is to consider 
its variation with respect to the gauge field configuration
for the following reason.  Let $Z$ be the partition function 
for an arbitrary (generically interacting) chiral theory on the 
lattice in the Ginsparg-Wilson formalism given by
\[
Z=\int\prod\ud\bar c_i \ud c_j e^{S[\bar c_i u_i,\, c_j v_j,\, U_\mu(x),\, O]}\,,
\]
where $O$ represents collectively the operators that appear in the
theory. If ${\chi_a}$ is a set of coordinates on the gauge field
configuration space $\mathbb U$, it has been proven in \cite{Poppitz:2007tu}
that
\[
\partial_a\ln Z=
\sum_i \left(\partial_a u^\dag_i \, u_i+v_i^\dag\,\partial_a v_i\right)
+\vev{\frac{\delta S}{\delta O}\partial_a O}\,,
\]
where $\partial_a\equiv\partial/\partial\chi_a$.
Apart from the usual contributions expected, given
by the last term, an extra piece
\[
j_a=\sum_i \left(\partial_a u^\dag_i \, u_i+v_i^\dag\,\partial_a v_i\right)
\]
emerges solely due to the variation of the eigenvectors $u_i$ 
and $v_i$. $j_a$ is often named the ``measure current''
since it captures completely the arbitrariness of 
the phase of $Z$ caused by the free choices of
$(u_i, v_i)$. Imagine that a different orthonormal frame 
$u_i'=\sum_j\mathcal U_{ij} u_j$ is adopted, then, $j_a$
shifts by a total derivative as
\[
j'_a=j_a+\Tr(\partial_a \mathcal U^\dag\, \mathcal U)
=j_a-\partial_a \ln \det U\,.
\]
This means $j_a$ defines a connection on the $U(1)$ bundle
over $\mathbb U$. Therefore, in the rest of the paper, we shall
call it more appropriately the ``measure connection''. 
It is a well-known mathematical fact that a smooth global section on the said bundle
exists only if a smooth connection can be defined.

Now comes one of the most peculiar properties of the Ginsparg-Wilson
formalism. Even though the measure connection $j_a$ is never unique, 
the curvature tensor it defines is in fact fully determined by the 
Ginsparg-Wilson operator, with no ambiguity or singularities whatsoever,
since
\begin{equation}\label{eq:curvature}
\mathcal F_{ab}=\partial_a j_b-\partial_b j_a=
\Tr\,P\commut{\partial_a P}{\partial_b P}\,.
\end{equation}
A quick way of proving this is by noticing that $P_+=\sum_i u_i u_i^\dag$.

It is well-known that the $2$-form
$\mathcal F=\mathcal F_{ab}\ud \chi^a\wedge\ud\chi^b$ integrated 
over a closed $2$-cycle is quantized.  Given the fact that 
$\mathcal F_{ab}$ is well defined globally on $\mathbb U$ without any singularities,
it must be true that
\begin{equation}\label{eq:master_integral}
\int_\tau \mathcal F=0
\end{equation}
on any $2$-cycles that can be continuously deformed into either a single point
or any lower dimensional cycles.  On the other hand, if there is a 
$2$-cycle $\tau$ in $\mathbb U$ over which
\[
\int_\tau\mathcal F>0\,,
\]
possible only if $\tau$ is non-contractible of course,
it indicates immediately that no smooth connection
$j_a$ on the entirety of $\mathbb U$ can ever be found, or, by the Stoke's
theorem, one must conclude $\int_\tau\mathcal F=\int_{\partial\tau} j=0$ since
$\partial\tau=0$. Hence, a non-zero integral of $\mathcal F$
over any $2$-cycle presents a true topological obstruction for
defining the associated chiral lattice gauge theories
in the Ginsparg-Wilson formalism.

Therefore, our task is to 
fully understand the topology of $\mathbb U[G]$, find all possible
non-contractible $2$-cycles in it, evaluate the integral \eqref{eq:master_integral}
over each, and make sure it vanishes always.
Provided this can can done, one still must further verify that the measure can be chosen
to stay invariant along the gauge obits in $\mathbb U[G]$
so that the partition functions are gauge invariant, and, along
other directions, while it can vary,
it must do so in the manner that it is expressible as a smooth gauge 
invariant local expressions so that
the arbitrariness for the choice of ``fermion measure'' amounts only
to an arbitrary local counter-term to be removed as the theory is renormalized
along its way toward the continuum limit.

All the steps above have been fully established for the abelian gauge theories,
and the program proved successful if and only if
the fermion content are chosen so
that the anomaly-free condition \eqref{eq:abelian_anomaly_cancellation}
is satisfied. For non-abelian gauge theories, only a perturbative
proof was known.  In this paper, we make one step toward the full solution for
the non-abelian gauge theories in $2$-d by considering the same questions
on a subset of $\mathbb U$, and we conjecture that
this consideration is sufficient.

\section{Solving the measure problem when gauge anomaly cancels}
\label{sec:anomaly_free}
It has been observed that the coincidence of the absence of topological 
obstruction explained above and the anomaly-free condition has a 
simple geometrical interpretation in $2$-d \cite{Neuberger:1998xn, Poppitz:2007tu}, focusing 
on an exceedingly restricted gauge configuration space.
An immediately natural step to take is to 
generalize those found for the abelian theories to the non-abelian ones.

Before we do so, a few words about the ``permissible gauge field
configuration space $\mathbb U[G]$ are in order. Naively, on
a $2$-d square lattice, one may think that the ``full'' gauge 
field configuration space is simply $G^{2N^2}$ since $2N^2$ is
the total number of the links. This turns out to be not true.
The space $G^{2N^2}$ is too large containing more cycles that
are irrelevant. Since ultimately, one is interested in taking
the continuum limit, the gauge field configurations should be 
confined within a subspace such that the field strength in each plaquette is bounded as
\[
|\ln\tr f(p)^2|<\epsilon
\]
so they do not produce overly many topological copies as
approaching the continuum.
Field configurations within such bound is what we
called the ``permissible configurations'', and they form the
space $\mathbb U[G]$ that is discussed throughout this paper. 

Being a subspace of $G^{2N^2}$, when $\epsilon$ is sufficiently
small, its topology is much simplified. This is essentially
the origin to equation \eqref{eq:abelian_u}. Most generally, 
if we assume $G^{2N^2}$ is non-singular in a small enough domain
surrounding the slice formed by the field configurations of
zero field strengths, denoted as $\mathbb U_0[G]$, near that
slice, it always takes the form of $\mathbb U_0[G]\times \mathbb F$ where
$\mathbb F$ is contractible. This is proven
for $G=U(1)$ but remain unchecked more generally. If
we take the position that this fact, or a somewhat modified version
of it, remains true for non-abelian $G$, within a range of
$\epsilon$, we find that considering the much simpler 
problem in $\mathbb U_0$ is justified, since it
is a well-known mathematical fact that a smooth connection
exists in the $U(1)$ bundle over $\mathbb U_0\times\mathbb F$ if and 
only if it does so in the same over $\mathbb U_0$ provided
that $\mathbb F$ is contractible. We hope to fully investigate
the space $\mathbb F$ closely in the future.

We remind the readers again that considerations on $\mathbb U_0$
is not just the zero-th order in a perturbative expansion since there
are important ``large'' Wilson-lines on such gauge backgrounds.

From now on, we focus only on $\mathbb U_0[G]$, and
the rest of the paper is devoted to first reviewing briefly $\int_\tau\mathcal F=0$
on the $T^2$ factor for $G=U(1)$, then showing to what extent one may reproduce the same
kind of calculation when $G$ is non-abelian and how the gauge anomaly cancellation 
formulae arise in a similar manner, and finally presenting the
detailed analysis of $\mathbb U_0[G]$ and proving equation \eqref{eq:non-abelian_u0}.

To be specific, the convariantized Ginsparg-Wilson operator is defined
as follows.  The gauge group $G$ is assumed to be
a semi-simple compact Lie group, and the fermions are in
some of its unitary representation. For each fermion species, we define
\begin{equation}\label{eq:X}
\begin{split}
X_{mn}=&\frac{1}{2}\sum_{\mu}\gamma_\mu\left(\delta_{m+\hat\mu,n}U_\mu(m)
		-\delta_{m, n+\hat\mu}U^\dag_\mu(n)\right)\\
	&+\frac{1}{2}\sum_\mu\left(\delta_{m, n+\hat\mu}U_\mu^\dag(n)
		+\delta_{m+\hat\mu, n}U_\mu(m)\right)-1\,,
\end{split}
\end{equation}
where $m$ and $n$ refers to lattice vertices, and then the Ginsparg-Wilson
operator
\[
D=1-\frac{X}{\sqrt{X X^\dag}} \gamma_5\,.
\]
The square-root of the Hermitian matrix $X X^\dag$ is defined by
taking the positive roots for each of the eigenvalues of $X X^\dag$.
Evidently $D$ satisfies the Ginsparg-Wilson relations \eqref{eq:ginsparg-wilson}. 
It follows that $\hat\gamma_5$ is given by
\[
\hat\gamma_5=\frac{X}{\sqrt{XX^\dag}}\,.
\]
They are covariant operators in the sense that under the gauge transformation
\[
U_\mu(m)\rightarrow \omega(m+\hat\mu)U_\mu(m)\omega(m)^{-1}\,,
\quad\textrm{for~}m\in\mathbb L\,,\; \mu=1, 2\,,
\]
they transform as
\[
D_{mn}\rightarrow \omega(m)D_{mn}\omega(n)^{-1}\,,\quad
\hat\gamma_{5\, mn}\rightarrow \omega(m)\hat\gamma_{5\, mn}\omega(n)^{-1}\,.
\]

\subsection{The abelian story: wrapping a torus on a sphere}
\label{sec:abelian}
For abelian group, the space $\mathbb U_0[U(1)]$ was known to take the form
of $T^2\times U(1)^{N^2-1}$, where the torus $T^2$ describes the two gauge
invariant measures given by the so-called nontrivial Wilson-lines across
the lattice:
\[
w_\mu=\prod_{s=0}^{N-1} U_\mu(x+s\hat\mu)\,.
\]
The choice of $x$ is irrelevant. Obviously $w_\mu=\exp\{i\theta_\mu\}$
where $\theta_\mu\in[0, 2\pi)$.  A typical field configuration corresponding
to a set of $(w_1, w_2)$ is given by
\begin{equation}\label{eq:abelian_homo_w}
U_\mu(x)=e^{i\theta_\mu/N}\,.
\end{equation}
Certainly, infinitely many other zero field strength configurations
corresponding to the same $w_\mu$ exist but they are all related to
each other by gauge transformations summarized by the $U(1)$ factors
of $\mathbb U_0$. Ignoring the gauge transformations,
to be justified in section \ref{sec:gauge_transformation},
the simple homogeneous field configuration is all we need to care about.

On translationally symmetric backgrounds, everything is most conveniently 
expressed in the momentum space, in which the operators are
block diagonal. On an $N\times N$ periodic lattice, 
momenta are periodic variables between $0$ and $\pi$ (as 
normalized in \cite{Poppitz:2007tu}) and discretized in units of $\pi/N$.
Setting $\theta_\mu=0$ at the moment,
the Ginsparg-Wilson operator $D$ given above turns into
\[
D_0(\vec p, \vec q)=\delta_{\vec p, \vec q}\, d(\vec p)
\]
in momentum space \cite{Giedt:2007qg, Poppitz:2007tu}, where
\[
d(\vec p)=\left(
\begin{array}{cc}
a(\vec p) & i c(\vec p)+ b(\vec p)\\
i c(\vec p)- b(\vec p) & a(\vec p)
\end{array}\right)
\]
and
\[
\begin{split}
a(\vec p)&\equiv 1-\frac{1-2 s(p_1)^2-2 s(p_2)^2}{v(\vec p)}\\
b(\vec p)&\equiv\frac{s(2p_2)}{v(\vec p)}\\
c(\vec p)&\equiv\frac{s(2p_1)}{v(\vec p)}\\
v(\vec p)&\equiv \sqrt{1+8 s(p_1)^2 s(p_2)^2}\\
s(x)&\equiv\sin x\,,\qquad c(x)\equiv \cos x\,.
\end{split}
\]
Turning on the homogeneous background \eqref{eq:abelian_homo_w}
is easy. Substituting 
$U_{\mu}(x)=\exp\{q\, \theta_\mu/2N\}$\footnote{Recall that,
specified for each flavor, the charge $q$ needs to be restored.}
in \eqref{eq:X} is evidently
equivalent to shifting the momentum variable
$p_\mu$ by a constant $q\theta_\mu/(2N)$. Therefore
\[
D(\vec p, \vec q, \vec \theta)=
D_0\left(\vec p+\frac{q \,\vec \theta}{2N}, \vec q+\frac{q\,\vec \theta}{2N}\right)\,.
\]
Now we can calculate $\mathcal F$ and its integral explicitly. Surely
the only interesting $2$-cycle on $T^2$ is $T^2$ itself and so the only
integral to check is $\int_{T^2}\mathcal F$.  

It turns out that even such a simple calculation one can be excused from.
What really matters here is that the operators are block diagonal in the
momentum space, so for each momentum mode, 
$D$, $P_+$, and $\hat\gamma_5$ are simple $2\times 2$ matrices.
Particularly, $\hat\gamma_5(\vec p, \vec \theta)$ 
is a $2\times 2$ Hermitian matrix satisfying $\hat\gamma_5^2=\mathbf 1$
and $\tr\hat\gamma_5=0$.  Any $2\times 2$ matrix of such kind can be represented by
\[
\hat\gamma_5(\vec p, \vec \theta)=\hat n(\vec p, \vec\theta) \cdot\vec\sigma\,,
\]
where $\hat n$ is a $3$-dimensional unit vector whose tip sits
on a $2$-sphere.  $\vec\sigma=(\sigma_1,\sigma_2, \sigma_3)$ are the Pauli matrices.
A few steps of calculation show
\[
\mathcal F_{\mu\nu} \ud \theta_1\wedge \ud \theta_2
=\Tr{P_+\commut{\ud P_+}{\ud P_+}}
=\frac{i}{2}\sum_{\vec p}\vec n\cdot(\ud \vec n\times\ud \vec n)\,.
\]
Each term in the above summation is a projected area form on
the unit sphere. So the $2$-form 
$\Sigma(\vec\theta)=\sum_{\vec p}\vec w(\vec p, \vec\theta)
	\cdot[\ud \vec w(\vec p, \vec\theta)\times\ud \vec w(\vec p, \vec\theta)]$
defines the same that is periodic in $\vec\theta$ because the left-hand side is.
Obviously
\[
\int_{T^2} \Sigma(\vec h)=2\pi n_w\,,
\]
where $n_w\in \mathbb Z$ is the wrapping-number representing the times
the unit sphere is wrapped over by $T^2$ determined by the
mapping $\hat n(\vec\theta)$. 
To find $n_w$, it suffices to investigate the mapping
at some particular point that is most convenient. In this case, it is around the
north pole on $S^2$ when $\hat n=(0, 0, 1)$. This is reached only by setting
$\vec \theta=0$ when $q=1$, so one immediately finds $n_w=1$.
For $q>1$, nearby each
point $\theta_\mu=2\pi i_\mu /q$, for $i_\mu=0, 1, \dots, q-1$, the operator $D$ 
appears identical and so we must find 
\[
n_w=q^2\,.
\] 
Considering multiple flavors with different charges
and the proper signs for fermions of either chirality,
we arrive at the anomaly-free formula
\[
\int_{T^2}\mathcal F=0\quad \textrm{if and only if}\quad \sum_i q_{L, i}^2=\sum_j q_{R,j}^2\,,
\]
as advertised.

\subsection{A first attempt to attack the non-abelian theories}
\label{sec:non-abelian}
Let us try to generalize the above result to non-abelian theories for
as much as we can.  For abelian groups, homogeneous field configurations 
automatically have zero field strength. The same is not true in the 
non-abelian case except
when the two Wilson-lines winding $\mathbb L$ in both directions commute.
More careful analysis is presented in Sec. \ref{sec:non-abelian}, but
right now, this prompts us to consider the simplest 
possible configurations given by
\begin{equation}\label{eq:abelianize}
U_\mu(x)=\exp\{i\theta_\mu t_\mu/N\}\,,\quad\mu=1,2\,,
\end{equation}
where $t_1$ and $t_2$ are two element in the Lie algebra $\mathfrak g$
for $G$ \emph{chosen to commute}.  This field configuration produces
the Wilson-lines $(w_1, w_2)=(\exp\{\theta_1 t_1\}, \exp\{\theta_2 t_2\})$ 
winding the lattice in both directions.  In other words, we choose to 
completely ignore the non-abelian
nature of $G$ and focus only on one of its maximal abelian subgroups. Such
a subgroup is the maximal torus of $G$, which we denote as $T_k$ in the following.
It is a $k$-dimensional torus where $k=\textrm{rank} G$.  Correspondingly
$t_1$ and $t_2$ are members of the Cartan subalgebra $\mathfrak c\subset\mathfrak g$.
Both the maximal tori and the Cartan subalgebra are not unique, but different
ones are isomorphic and choosing an arbitrary pair leads to equivalent results.  

We note that even with the condition that $w_1$ and $w_2$ commute, 
field configuration \eqref{eq:non-abelian_homo_w} is overly restrictive since
both $t_\mu$, while commuting, can in principle be $\theta_\mu$ dependent,
corresponding to letting $w_\mu$ wander from one maximal torus to another
as $\theta_\mu$ vary.  Such a complication will be shown to be removable
by gauge transformations.

On the background \eqref{eq:abelianize},
all the work done in the previous section is easily duplicated.  
Given that $t_1$ and $t_1$ commute, in an appropriate basis,
they can be simultaneously diagonalized. Consequently, $U_\mu(x)$ are diagonal matrices
with respect to the group indices as
\begin{equation}\label{eq:non-abelian_homo_w}
U_\mu(\theta_\mu, x)=\left(
\begin{array}{cccc}
e^{i \theta_\mu/N t_\mu^1} &&&\\
& e^{i \theta_\mu/N t_\mu^2} &&\\
&&\ddots&\\
&&&
e^{i \theta_\mu/N t_\mu^d}
\end{array}\right)\,,
\end{equation}
where $t_\mu^i$ are the $i$-th diagonal entry of $t_\mu$ and $d$ is the dimension
of the representation. Notice that the assumption that $t_\mu$ are constants
plays a vital role here, since if they do vary with $\theta_\mu$, even though
$t_1(\theta_\mu)$ and $t_2(\theta_\mu)$ are simultaneously diagonalizable at 
each fixed $\theta_\mu$, generically the diagonal form can not
be kept as $\theta_\mu$ vary.

Evidently, entry by entry, inserting a factor \eqref{eq:non-abelian_homo_w}
in \eqref{eq:X} amounts to shifting the momentum $\vec p$ by a constant
of $t^i_a\theta_\mu$ just like in the abelian case.
It is as if we have a $U(1)$ gauge field with $d$-multiple of fermion
species, each with an effective charge $t_\mu^i$. The only difference here is
that a single fermion species seemingly has different effective charges with respect to
the gauge field along different dimensions, something impossible in the genuine
$U(1)$ theory. This minor discrepancy does not affect the calculation much.
So, a single fermion species contributes a term to $n_w$ as
\[
n_w \textrm{~by each fermion multiplet}=
\sum_i^d \, t^i_1 t^i_2=\tr \, t_1 t_2\,.
\]
and the vanishing of the total wrapping number is precisely
given by equation \eqref{eq:non-abelian_anomaly_cancellation}.
Once again, this result is derived with the assumption that
$t_1$ and $t_2$ lie in one copy of the Cartan subalgebra of $\mathfrak g$,
but this is sufficient to prove the same holds for all $t\in \mathfrak g$.

\subsection{The justification for omitting the gauge transformations}
\label{sec:gauge_transformation}
We fill in one remaining gap in the above reasoning here:
that directions in $\mathbb U_0$ corresponding to gauge
transformations can be ignored.

Let $\lambda_a$ be a set of coordinates parameterizing
the gauge group $G$. A general gauge transformation is specified
by the functions $\lambda_a(x)$ where $x$ is the vertex
coordinate in $\mathbb L$, so a gauge transformation 
labeled by  $\lambda_a(x)$ is generated by
the group valued function $\omega(x)=g(\lambda_a(x))\in G$.
To compress the notations, we may also write the coordinates as
$\lambda_{a, x}$. Suppose the space $\mathbb U_0$ is endowed
with a set of coordinates $(w_a, \lambda_{b, x})$, where
$w_a$ specify the directions ``perpendicular'' to the gauge
orbits and $\lambda_{b, x}$ parameterize each gauge orbit or the slices
in $\mathbb U_0$ generated by gauge transformations
\footnote{More rigorously, there are usually constraints among
$\lambda_{b, x}$ but this does not matter.}. When no confusion is
caused, we suppress the subscript for $w_a$.

We assume that the slice in $\mathbb U_0$
at $\lambda_{a, x}=0$ for all $(x, a)$ corresponds 
to the gauge choice that the gauge field configuration
is translationally symmetric as given in \eqref{eq:abelian_homo_w}
or \eqref{eq:non-abelian_homo_w}, and denote it
as $\mathbb W_0$.

The goal is to prove that given a smooth connection $j^\circ_w$ on the
subspace $\mathbb W_0$, one can always extend it
to $\mathbb U_0$, including the definition for the new components
$j_{\lambda_{a,x}}$, which enjoys the properties:\\
\hspace*{10 mm}
i) \emph{Both $j_w$ and $j_{\lambda_{a,x}}$ are smooth on $\mathbb U_0$, and}\\
\hspace*{10 mm}
ii) \emph{$j_w$ and $j_{\lambda_{a,x}}$ are gauge invariant, i.e. they are
$\lambda_{a,x}$-independent.}

Let the smooth connection on $\mathbb W_0$ be given by
the set of eigenvectors $u^\circ_i(w; x)$ and $v^\circ_i(x)$
\footnote{A noteworthy fact is that these vectors 
are \emph{never} smooth functions of the gauge field configurations
even when the anomaly-free condition satisfied, a fact that
we can not elaborate on here. But a quick argument for it is
that should they be chosen so, one may keep one fermion
species only and drop all the rest and nothing would prevent 
the vectors $u_i$ to remain smooth, consequently giving
rise to a smooth connection $j_a$ even with anomalous field content. 
Only the connection $j_a$ or the corresponding
phase function of $Z$ might be smooth.}, so on $\mathbb W_0$
\[
j^\circ_w=\sum_i \partial_w u^{\circ\dag}_i(w'; x)\, u^\circ_i(w'; x).
\]
A superscript ``$\circ$'' indicates quantities evaluated on the slice
$\mathbb W_0$. Clearly, $v^\circ_i$ can be chosen to be $w$-independent.
Moving away from the point $\lambda_{a, x}=0$, the gauge field
changes by gauge transformations generated by $\omega(x)=g(\lambda_{a, x})$.
Since the operator $\hat\gamma_5$ is covariant, 
the new eigenvectors are easy to find and we may choose
\[
u_i(w, \lambda_{x', a}; x)=g[\lambda_a(x)]u^\circ_i(w; x)\,,\quad
v_i(w, \lambda_{x', a}; x)=g[\lambda_a(x)]v^\circ_i(w; x)\,.
\]
Certainly, this is not a unique choice but happens to be the one we
would use. Obviously, with this basis, $j_w=j_w^\circ$ on the 
entire $\mathbb U_0$, and so is both smooth and gauge invariant.

The new components for the measure connection along the directions
of the gauge orbits are given by
\[
\begin{split}
j_{\lambda_{a, x}}=&\sum_i\left\{ u_i^{\circ \dag}
\left[\partial_{\lambda_{a,x}}g (\lambda_a(x))^\dag\right]
\, g(\lambda_a(x))u_i^\circ\right.\\
&\qquad+\left.v_i^{\circ \dag} g(\lambda_a(x))^\dag\, 
\left[\partial_{\lambda_{a,x}}g (\lambda_a(x))\right]
v^\circ_i\right\}\,.
\end{split}
\]
Using the fact that
\[
\sum_i (u^\circ_i u_i^{\circ\dag} -v^\circ_i v^{\circ\dag}_i )=
\hat P^\circ_{+ xx}-P^\circ_{+ xx}
=\frac{1}{2}(\hat\gamma^\circ_{5xx}+\gamma_{5xx})
\]
we find
\[
j_{\lambda_{a, x}}=\frac{1}{2}\tr
\left[\partial_{\lambda_{a,x}}g (\lambda_a(x))^\dag
\, g(\lambda_a(x))\left(\hat\gamma^\circ_{5xx}+\gamma_{5xx}\right)\right]\,.\\
\]
Here ``$\tr$'' denotes the trace over both the fermionic and group
indices at a fixed lattice position $x$.

When the gauge group is abelian, $\partial_{\lambda_{a,x}}g^\dag\, g=q$ 
is a $c$-number and can be pulled out
out the trace.  Using the fact that gauge field configurations on 
$\mathbb W_0$ are homogeneous with zero field strength, we find
\[
\tr{\left(\hat \gamma^\circ_{5 xx}+\gamma_{5 xx}\right)}
=\frac{1}{N^2}\Tr(\hat \gamma^\circ_5+\gamma_5)=0.
\]
Therefore, $j_{\lambda_{a,x}}=0$, which obviously satisfies both properties required.

When the group $G$ is non-abelian, we would have to resort to the
simplification that the link fields on $\mathbb W_0$ are chosen
to sit within a maximal abelian subgroup of $G$, and by choosing the basis
appropriately, the link fields are simple diagonal matrices as given in 
equation \eqref{eq:non-abelian_homo_w}.
On such backgrounds, with respect to
the group indices $\hat\gamma^\circ_{5 xx}$ is diagonal and each
diagonal entry must be identical to the same operator in the abelian theory
with only the fermion charge $q$ replaced by the diagonal entries
$t_\mu^i$.

In the meanwhile, $t=(\partial_{\lambda_{a, x}}) g^{-1} g$ is an
Lie algebra element in $\mathfrak g$. Now that both $\hat\gamma_{5xx}$
and $\gamma_{5xx}$ are diagonal with respect to the group indices,
only the diagonal part of $t$ matters once the trace is
taken.  Once again, we are ready to recycle the known results from the abelian theories
since we have essentially expressed the connection $j_{\lambda_{a, x}}$
as a sum of many copies, each obtained effectively from an abelian theory.
It follows that $j_{\lambda_{a,x}}=0$.
 
Therefore, it is sufficient to construct a smooth measure connection on
$\mathbb W_0$, if we can show that the space $\mathbb U_0[G]$ can be parameterized
as assumed in the beginning of this subsection even for non-abelian groups. 
We turn to this topic next.

\section{The field configuration space of zero field strength for non-abelian $G$}
\label{sec:main}
In this section, we shall investigate the gauge field configuration space of
zero field strength for an arbitrary non-abelian
gauge group $G$, denoted as $\mathbb{U}_0(G)$, and prove the main
result mentioned around \eqref{eq:non-abelian_u0}.  A more explicit
toy example for $G=SU(2)$ is presented in the end for entertaining.

\subsection{The space $\mathbb U_0[G]$}
\label{sec:u_0}
We aim at describing the space $\mathbb U_0[G]$ for a non-abelian gauge group
in a similar manner as in the abelian case.
The objects that we are interested in for any gauge field configuration
are the Wilson-lines along some path in $\mathbb L$ formed by a sequence of
consecutive links $l_i(x_i)$, $i=1, 2, \dots, s$, defined by the path-ordered 
product
\[
w_{\mathcal C}=U_{\mu_s}(x_s)U_{\mu_{s-1}}(x_{s-1})\dots U_{\mu_2}(x_2) U_{\mu_1}(x_1)\,,
\]
which is a group element in $G$.  When $x_s+\hat\mu_s=x_1$, 
$\mathcal C$ forms a closed loop and the corresponding Wilson-line is also
referred to as the Wilson-loop. The minimal
Wilson-loops that can be formed on a square lattice are those along the 
the four links surrounding each single plaquette. We refer to them as the 
``field strength'', denoted as $f(p)$ for the plaquette $p$.

We should emphasize that we have defined the Wilson-lines without taking the trace 
for the convenience of discussion, and so it depends on the choice of the starting
point for non-abelian groups even if $\mathcal C$ is closed. More precisely,
we should denote it as $w_{x, \mathcal C}$, referred to as the Wilson-loop based at 
the point $x$. Similarly, the ``field-strength'' should be more properly
denoted as $f(x, p)$. Evidently, two Wilson-loops along the same closed path 
$\mathcal C$ but based at different points, say at $x$ and $x'$ respectively,
differ by a group conjugation. If the two Wilson-lines are $w_1$ and $w_2$, 
$w_1=g w_2 g^{-1}$, where $g$ is the Wilson-line along the section of 
$\mathcal C$ connecting $x'$ to $x$.

Nor are the Wilson-lines defined in this manner gauge invariant. Upon a gauge
transformation generated by $\omega(x)$, evidently, $w_{x, \mathcal C}$ 
transforms by conjugation as well as 
\[
w_{x\mathcal C}\rightarrow \omega(x) w_{x\mathcal C}\omega(x)^{-1}\,.
\]

Consequently, that $w_{x,\mathcal C}=\mathbf 1$ is base point
and gauge choice independent.  
The gauge field configuration is said to have zero field strength if
\[
f(p)=\mathbf 1\quad\textrm{for~}\forall p\in \mathbb L\,.
\]
And the space of all such field configurations has been denoted as $\mathbb U_0$.

We prove the following theorem:
\begin{thm}\label{thm:non-abelian_u0}
On a $2$-d $N\times N$ square lattice with periodic boundary conditions, the 
space $\mathbb U_0=S(G)\times G^{N^2-1}$. The factor $G^{N^2-1}$ corresponds
to gauge transformations and $S(G)=\{(g_1, g_2)|g_1 g_2=g_2 g_1, g_1, g_2\in G\}$,
which can be considered as $T_k^2$ foliated by the conjugacy classes of
$G$. Each conjugacy class is a slice generated by acting
on points in $T_k^2$ by gauge
transformations, and is therefore a gauge orbit.
Here $T_k$ is any one of the maximal tori in $G$.
\end{thm}

The proof is fairly straightforward. The foundation of all 
is the following counterpart of the Stoke's theorem in the non-abelian case. Consider
a sub-lattice $\mathbb D\subset\mathbb L$ consisting of a collection of
plaquettes, and its boundary $\mathcal C=\partial \mathbb D$, which is
always a closed loop, the Wilson-loop $w_{\mathcal C}$ 
is determined by the ``field strength'' of each plaquette in $\mathbb D$ as
\[
W_{\mathcal C}=\prod_{p\in \mathbb D} f(p).
\]
\emph{when the gauge group is Abelian}. Such simple
formula does not exist for non-abelian groups. 
For an example, the Wilson-loop around the three adjacent 
plaquettes $p_1$, $p_2$, and $p_3$ is in general not expressible as the product of $f(p_1)$,
$f(p_2)$, and $f(p_3)$ in any order no matter how the base points are chosen.
However, a modified ``Stoke's theorem'' does exist and it says
\begin{lem}\label{lem:non-abelian_stokes}
On any $2$-d square lattice, let $\mathbb D\subset \mathbb L$ be a connected
sub-lattice, $\mathcal C=\partial\mathbb D$, and $x\in \mathcal C$ be an
arbitrary point on $\mathcal C$,
\[
w_{x, \mathcal C}=\mathcal P\prod_{p\in \mathbb D} c\, f(x_p, p) \, c^{-1}\,.
\]
Here $\mathcal P\prod$ denotes a path-ordered product. The actual order is not
particularly important here other than that it exists.
$c\in G$ is the Wilson-line along some path that connects
the base point $x$ to that of each plaquette $p$, i.e. $x_p$.
The choices for $c_p$'s also depend on the path-ordering and are usually
not arbitrary and mutually dependent.
\end{lem}
The theorem can be proven by induction, \footnote{See \cite{fishbane, broda} for
the counter part of it in the continuum.} as detailed in
Appendix \ref{app:stokes}.  For now, we use it to prove a simple
but powerful fact that is 
\begin{lem}\label{lem:trivial_loop_w}
On an arbitrary square lattice, when  the gauge field configurations have
zero strength, i.e. $f(p)=\mathbf 1$ for $\forall p\in \mathbb L$,
the Wilson-line around any closed loop $\mathcal C\subset\mathbb L$
that belongs to the trivial homology class, i.e. it forms the boundary of
a sub-lattice $\mathbb D$ in $\mathbb L$, is trivial:
\[
w_{x, \mathcal C}=\mathbf 1\,.
\]
\end{lem}
This follows from the Lemma just above since
$c\mathbf 1 c^{-1}=\mathbf 1$.

While the Wilson-lines around trivial loops are trivial, we should
examine what might be concluded for those around the non-trivial
circles in $\mathbb L$, i.e. those that can not be considered
as the boundary of any subsets. 
In the abelian case, it is fairly obvious that
on the zero field strength background, the basic
data consists of only two elements, $(w_1, w_2)$,
each corresponding to the Wilson-line
along the cycle that winds the periodic lattice once in either direction.
Around the more complicated loops, the result depends only
on how many times it winds around the lattice in either
dimension and can be expressed as a product of $w_\mu$.

\begin{figure}[ht!]
\begin{center}
\includegraphics[width=0.75 \textwidth]{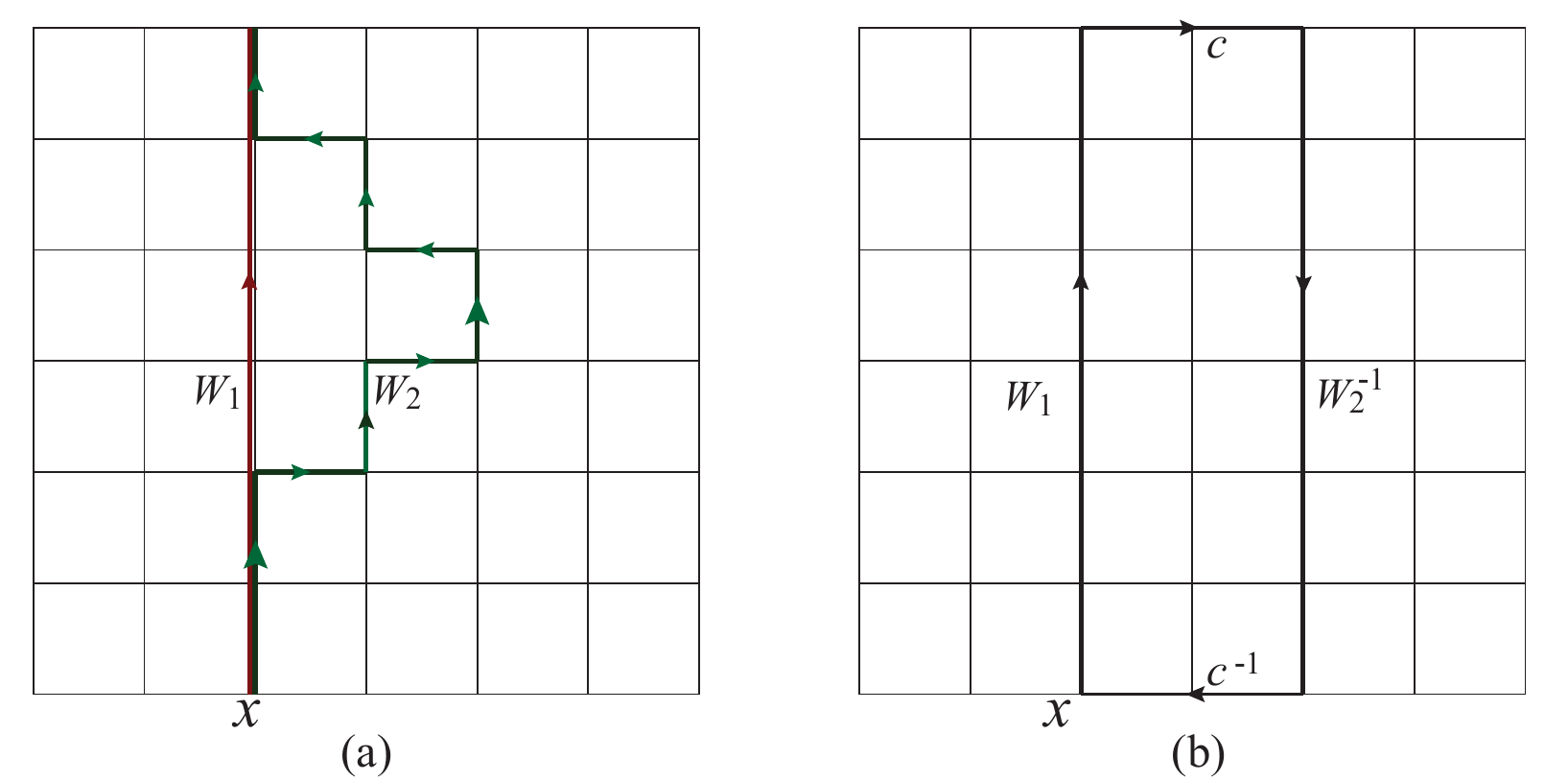}
\caption{\label{fig:twowilsons}}
\end{center}
\end{figure}

The situation is somewhat more involved in the non-abelian case. 
First, we notice that Wilson-loops crossing the lattice in either
direction once depend only on its starting point, i.e. 
the two loops as indicated in figure \ref{fig:twowilsons} (a) that share
the same base point but wander about in $\mathbb L$ along different paths 
while crossing the lattice have equal Wilson-loop. To see this is true,
just notice that one may reverse the order of the second path
and connect it to the first so that $\mathcal C'=\mathcal C_1\circ \mathcal C_2^{-1}$ 
forms a closed loop that is the boundary of some sub-lattice.
By Lemma \ref{lem:trivial_loop_w}, we find
$w_1 w_2^{-1}=\mathbf 1$ and so $w_1=w_2$. 

To further reduce the redundancy, consider two Wilson-lines crossing
the lattice in the same direction but starting from two different locations
as shown in figure \ref{fig:twowilsons} (b).  If we reverse the order of the second 
loop and connect it to the first by
inserting two identical paths in opposite directions in either end of it 
as shown in the figure, we make a closed loop that forms the boundary of a sub-lattice again.
By the same reasoning, we find $w_2 =c w_1 c^{-1}$ where
$c$ is the Wilson-line along the path that connects the base points for
$w_1$ and $w_2$. Recall that the Wilson-lines transform by conjugations 
upon gauge transformations. By a gauge transformation generated by
$\omega(x_2)=c^{-1}$, where $x_2$ is the starting point
$w_2$, and $\mathbf 1$ otherwise, the two Wilson-lines can be made equal.
Therefore, even for non-abelian groups, as long as
the field strengths vanish, up to gauge transformations,
there are only two ``large Wilson-lines''.
To be specific, we may choose $(w_1, w_2)$ as
\begin{equation}\label{eq:non-abelian_w}
w_\mu=\prod_{i=0}^{N-1} U_\mu(i \hat\mu)\,.
\end{equation}

\begin{figure}[ht!]
\begin{center}
\includegraphics[width=0.85 \textwidth]{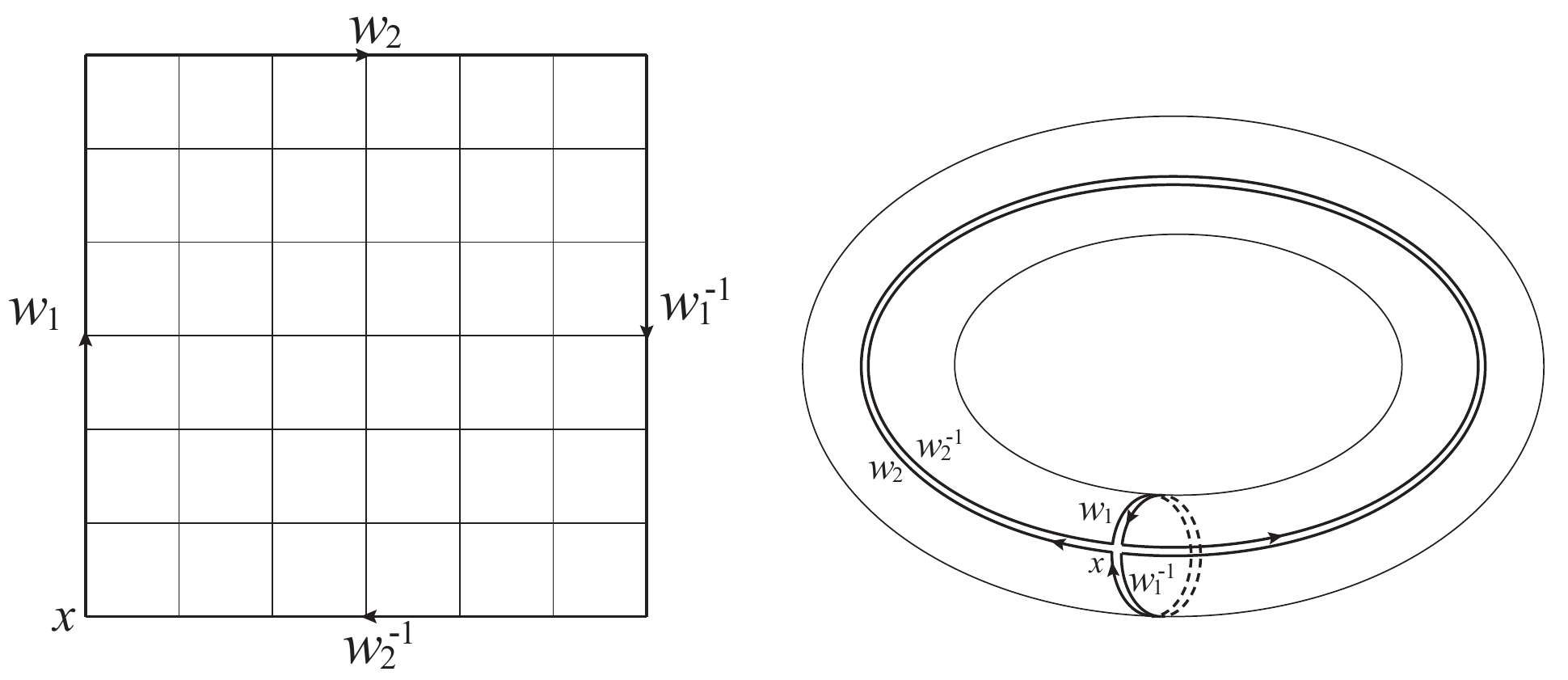}
\caption{\label{fig:perpwilsons}}
\end{center}
\end{figure}

Let us consider $(w_1, w_2)$ more closely.
Consider the closed loop shown in figure \ref{fig:perpwilsons}.
Connecting the two perpendicular cycles one after the other
twice in opposite directions forms a closed loop that belongs to the trivial
homology class. The left panel shows
the path represented by a lattice on a plane endowed with
periodic boundary conditions and the right panel shows how it would 
look like on the torus. Again by Lemma \ref{lem:trivial_loop_w},
we have
\[
w_1 w_2 w_1^{-1} w_2^{-1}=\mathbf 1,
\]
which is equivalent to $w_1 w_2=w_2 w_1$. So the two perpendicular Wilson-lines
must commute.
Finally, let us exploit the gauge freedom one more time. The two
Wilson-lines as defined in \eqref{eq:non-abelian_w} are not gauge
invariant. But their gauge transformations are tied together such 
that any gauge transformation conjugate both of them by a same group
element simultaneously. It is well-known that, in any compact Lie group, two commuting elements
always lie in a common maximal torus in $G$. There are infinitely
many maximal tori, but they form a single conjugation class in the sense
that any group conjugation just ``rotates'' one to another and conversely
two different tori can always be converted from one to the other by some conjugation.
Therefore, by exploiting this residual gauge freedom, we can always restrict
$(w_1, w_2)$ in some pre-chosen maximal torus $T_k$, and hence
almost prove theorem \ref{thm:non-abelian_u0}.

The remaining steps are standard. Let us embed the finite
lattice $\mathbb L$ in an infinitely large one $\mathbb L'$ 
that extends in both directions indefinitely. We think of $\mathbb L$ 
as the part bounded by $x\in[0, N-1]$ and $y\in[0, N-1]$
with periodic conditions imposed so that $U_1(i, 0)=U_1(i, N)$ and
$U_2(0, i)=U_2(N, i)$ for $i=0, 1, 2,\dots,N-1$.  On $\mathbb L'$, any
zero field strength gauge field configuration is $1$--$1$
correspondent to a group valued function $\omega(m)$ with the constraint
$\omega(0)=\mathbf 1$. Given any such $\omega$, the link field
\begin{equation}\label{eq:pure_gauge}
U_{\mu}(x)=\omega(x+\hat\mu)\omega(x)^{-1}\,,
\end{equation}
evidently has zero field strengths. On the other hand, given any link field $U_\mu(x)$
with zero field strengths, one may construct the function
$\omega(x)$ by picking a path that connects
$(0, 0)$ and $x$ formed by the sequence of links $l_{\mu_i}(x_i), i=1, 2, \dots, s$, where
$x_1=(0,0)$ and $x=x_s+\hat \mu_s$, and assigning
\[
\omega(x)=U_{\mu_s}(x_s)U_{\mu_{s-1}}(x_{s-1})\dots U_{\mu_1}(x_1)\,.
\]
The construction is consistent since if one reaches from $(0,0)$ to
$x$ along two different paths and end up with two definitions, $\omega_1$ and $\omega_2$,
$\omega_1=\omega_2$ followed by Lemma~\ref{lem:trivial_loop_w}. 

In general, $\omega(x)$ does not exist on $\mathbb L$ 
unless $\prod_{i=0}^{N-1} U_\mu(x+i\hat\mu)=\mathbf 1$,
or $\omega(x)$ thus constructed can not be periodic, and
nor is its definition path-independent.
Still, nothing prevents us from using \eqref{eq:pure_gauge}
as a bookkeeping device for $U_\mu(x)$ if we have the embedding picture
in mind.  In this language, $w_1=\omega(N,0)$ and $w_2=\omega(0, N)$.
To impose periodic boundary conditions on the link field demands
\[
\begin{split}
\omega(N, i)\omega(N, 0)^{-1}=\omega(0, i)\,,\quad&
\omega(i,N)\omega(0, N)^{-1}=\omega(i, 0)\,,\\
\omega(N, 0)=\omega(N, N)\omega(N, 0)^{-1}\,,\quad&
\omega(N, N)\omega(0, N)^{-1}=\omega(N, 0)\,,\\
&i=1,2\dots,N-1\,.
\end{split}
\]
The second line above is equivalent to $w_1 w_2=w_2 w_1$. Given these constraints,
only those field $\omega(x, y)$ for $x, y\in[1,N-1]$ are truly free parameters
contributing of factor of $G^{N^2-1}$.  $\omega(N, 0)$ and $\omega(0, N)$ are free
subject to the condition that they commute. Therefore, we find
\[
\mathbb U_0[G]=S(G)\times G^{N^2-1}\,.
\]
Obviously, the factors $G^{N^2-1}$ correspond to free gauge transformations. The space
$S(G)$ can always be parameterized by two group element in the maximal torus
foliated by the conjugacy classes. As discussed already above, those conjugacy classes
are simply the gauge orbits that connect the pair $(w_1, w_2)$ to those located in
different copies of the maximal tori of $G$.

\subsection{The $SU(2)$ example}
\label{sec:su2}
Just for fun, let us investigate the space $\mathbb U_0[G]$ and the nontrivial
$2$-cycles on it for the case $G=SU(2)$ in a bit more detail, where
everything is readily visualized.
For simplicity, let us assume that the fermions are in the fundamental representation.
The group $SU(2)$ is a $3$-sphere, and an arbitrary $g\in SU(2)$ can be expressed
as
\[
g=\cos\theta\mathcal \mathbf +\hat n\cdot\vec\sigma\sin\theta\,,
\]
where $\hat n$ is a $3$-dimensional unit vector 
and $\vec\sigma=(\sigma_1, \sigma_2, \sigma_3)$ are Pauli matrices.
$\theta$ and $\hat n$ together form the spherical coordinates of the $S^3$.

The maximal tori of $SU(2)$ are the great circles passing through
both north and south poles, or the pairs of meridians that are
$180^\circ$ apart in longitudes, and any 
two elements in $SU(2)$ that commute must be located in one of those pairs.
The conjugation classes are the spheres of equal latitude, and they
act on the maximal tori by rotating them about the $z$-axis.

\begin{figure}[ht!]
\begin{center}
\includegraphics[width=0.75 \textwidth]{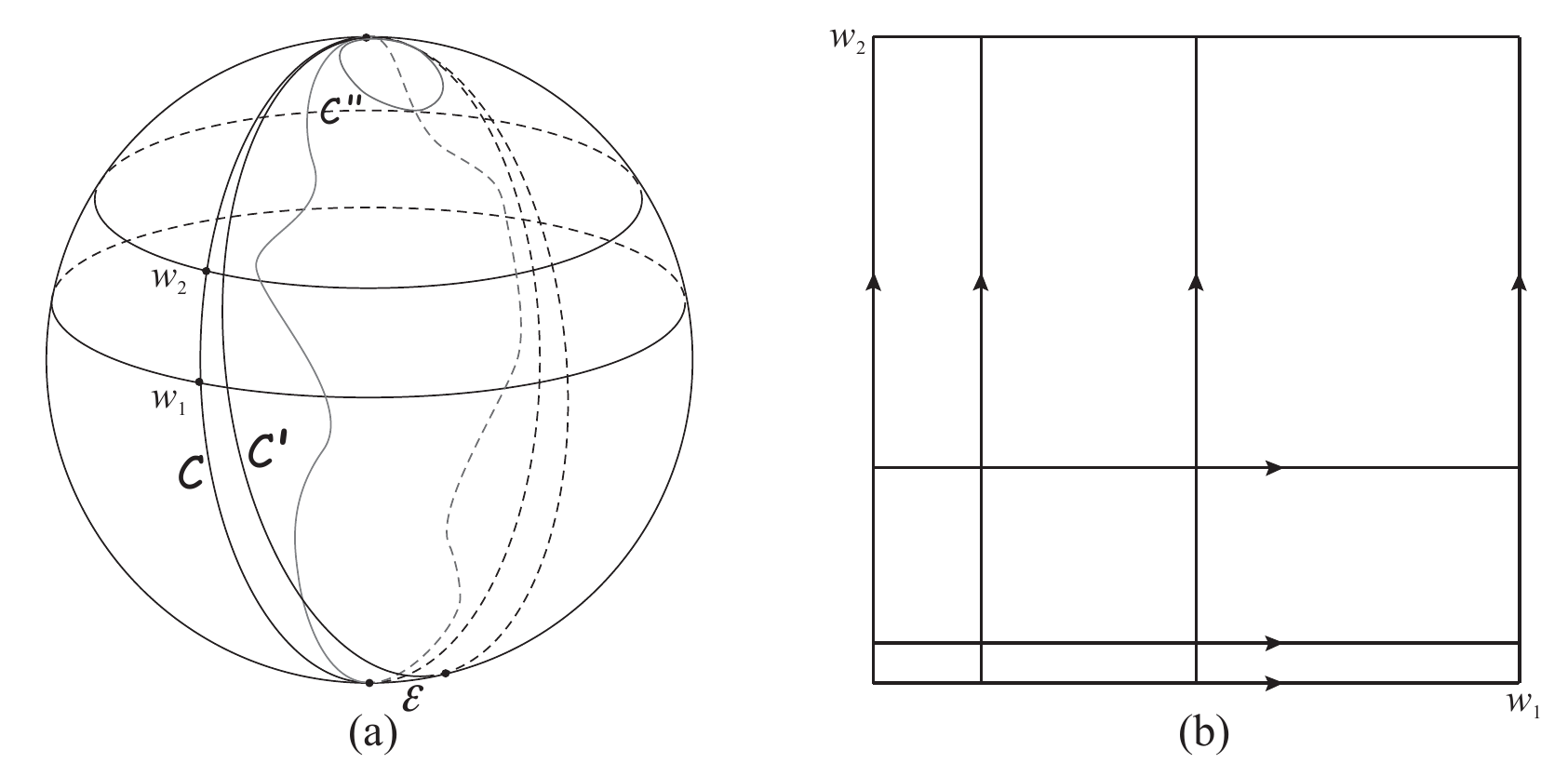}
\caption{\label{fig:su2}}
\end{center}
\end{figure}

Apart from the gauge transformations, $\mathbb U_0[SU(2)]$ is almost
identical to $\mathbb U_0[U(1)]\sim T^2$.
Choose one maximal torus as an representative, say the one formed
by the meridians at $0^\circ$ and $180^\circ$ longitude as shown
in figure \ref{fig:su2} (a). Let us call these pair of meridians
the standard meridians below.
The ``$T^2$ part'' is described by the two commuting Wilson-lines
$(w_1, w_2)$ that live on the standard meridian. 
We can parameterize their position by $(\theta_1, \theta_2)$,
which are nothing other than the latitudes of $(w_1, w_2)$
(or $2\pi$ minus that when they are on the meridian
at $180^\circ$).  

The conjugacy classes of $SU(2)$ are the
spheres at fixed latitudes. In figure \ref{fig:su2} (a),
they are depicted as circles with one dimension suppressed.
Each sphere is a gauge orbit representing the freedom,
by gauge transformations, of rotating the standard meridian about
the $z$-axis arbitrarily.  More precisely, the gauge orbits are
half of those spheres since any great circle intersects them 
twice as determined by the order of the Weyl group.

So, as $(w_1, w_2)$ wind around the space $\mathbb U_0$, they
appear to be winding around the $S^3$ freely subject only to
the constraint that they always lie in a common 
meridian (or the two of $180^\circ$ apart) at all time. 
In general, their trajectories
can be fairly complicated, curved and wiggled any way
they like, but by using the gauge transformations,
they can be ``straightened'' and rotated about the $z$-axis so
to coincide with the standard meridians without any obstructions.

Because $SU(2)\cong S^3$ and both fundamental groups $\pi_1$ and $\pi_2$ of
$S^3$ are trivial, one may suspect that it is possible to use the
gauge transformations to deform the trajectories of
$(w_1, w_2)$ away from the standard meridians such that the corresponding $2$-cycles
in $\mathbb U_0$ are always contractible to a single point, in 
which case $\int_\tau\mathcal F=0$. Our calculation in
Sec. \ref{sec:non-abelian} is sufficient 
to prove this false and the said integral can be
nonzero.  This tells that some $2$-cycles can not be continuously deformed
into a point, and it is still fun to think about pictorially
where the topological obstruction comes about.
Let us consider the $2$-cycle on $\mathbb U_0$ described by 
the coordinates $\theta_1, \theta_2\in [0, 2\pi)$ mentioned above.
We illustrate it in figure \ref{fig:su2} (b) by a square 
with $\theta_\mu$ being its coordinates.
Each horizontal line in this square represents a $1$-cycle describing $w_1$ winding
around the standard meridians once with $w_2$ kept fixed. Similarly,
each vertical lines represents the same with $w_1$ and $w_2$ interchanged.
Imagine we attempt to contract this $2$-cycle, utilizing the gauge freedoms.
Certainly, any small deformation to the $2$-cycles leads to a small
deformation to the said two kinds of $1$-cycles on $S^3$ represented by
the straight lines in \ref{fig:su2} (b). 
Somewhere very close to the horizontal axis, the horizontal lines in \ref{fig:su2} (b), after 
a small deformation, may represent a slightly deformed trajectory for $w_1$ on
$S^3$ as illustrated by the circle $\mathcal C'$ in figure \ref{fig:su2} (a).
Instead of cycling along the standard meridians, now it
runs around a slightly shifted circle
whose lowest point is away from the south pole by a tiny
distance $\varepsilon$. This trajectory is of a very good prospect for 
being continuously shrunk to a point near the north pole,
similar to the circle $\mathcal C^\dprime$ shown in the same figure.

However, such a deformation is not continuous for the
$2$-cycle.  As $w_1$ winds around, before the deformation, its
longitude stays fixed at either $0^\circ$ or $180^\circ$.
It does jump at one point, but evidently this is merely a coordinate
singularity.  After the small deformation, however, 
along the circle $\mathcal C'$, its longitude varies smoothly from $0^\circ$ toward $180^\circ$ 
passing by all values in between. At its lowest position,
it becomes $90^\circ$. This is inevitable the moment
one lifts the circle $\mathcal C$ off the south-pole
by however a small distance. Since $w_2$ is constrained to stay
at the same longitude with $w_1$ at all time, now the vertical line at the middle
part of the square in figure \ref{fig:su2} (b) can only represent
a $1$-cycle where $w_2$ winds around the
meridians at $\pm90^\circ$ longitude with $w_1$ kept at its 
lowest position on $\mathcal C'$.  This is a
big jump from the cycle represented by exactly the same
line when $\varepsilon=0$ no matter
how small $\varepsilon$ is. Therefore, the moment one 
attempts to lift the circle $\mathcal C$ off the south pole,
a discontinuous operation to the $2$-cycle is required.

\section{Conclusion and discussions}\label{sec:conclusion}
We found that if we restrict our attention to the zero field strength configurations,
the method that proves the existence of a smooth fermion measure for
the abelian theories applies straightforwardly to the non-abelian ones 
on $2$-d lattices. This is by no means surprising since
in $2$-d, essentially only the abelian subgroups participate in the gauge
anomalies. The detailed analysis showed the intuition was correct.

For a full proof, we must extend our work to include non-zero field strengths,
which requires significant amount of more technical work and we defer this
analysis to the future.
But it is quite conceivable that the main conclusion remains to hold. As mentioned
already, given the assumption that the full space $\mathbb U[G]$ is 
smooth near $\mathbb U_0[G]$, for sufficiently small $\epsilon$, the permissible
configuration space is expected to be given by $\mathbb U_0[G]\times\mathbb F$
where the second factor is contractible so that any smooth connection found
for zero field strength configurations extends automatically
to the full space easily. However, for finite $\epsilon$, this argument
fails to apply rigorously, and we hope to report more on this aspect soon.

Physically, of course, a much more interesting question is whether this
method can be applied to $4$-d. Even for the abelian theories, the problem is much harder.
As shown in \cite{Neuberger:1998xn}, it appears no longer sufficient to produce the needed 
anomaly cancellation formula without considering some nontrivial gauge field
background that contains monopole, although those considerations can be
motivated by the $2$-d analysis. The major difficulty is the topology of
$\mathbb U[G]$ becomes a lot more complicated, and we expect it is even
more so when $G$ is non-abelian. However, hints from the known
proofs for abelian theories suggest strongly that
most complicated topologies are irrelevant for which the integral $\int_\tau\mathcal F$ always
vanishes. It is possible that more intelligent study
may help to direct the attention to the actually cycles that
matter more straightforwardly.

Even when all the said work is done, and a ``good fermion measure'' is known
to exist for all chiral gauge theories when anomaly-free conditions
satisfied, there remains the challenge of actually implementing it on a computer
so real simulations can be done. In principle, the fermion measure can be
constructed directly, but the difficulty of doing so grows
as the size of the lattice and becomes essentially impractical 
\cite{Kadoh:2007wz, Kadoh:2007xb}. Methods
suggested in \cite{Eichten:1985ft,Giedt:2007qg} might be more feasible, 
where one starts with vector-like
theories and, by using the peculiar phases existing only on the lattices,
decouple half of the spectrum in the continuum limit such that a chiral
theories emerges automatically, bypassing all the explicit constructions.
The problem of this strategy is that in general it is very hard to prove
for a clear decoupling, particularly on non-trivial gauge field backgrounds.
In the meanwhile, one may also worry about the unitary and locality of
the emergent theory. However, earlier simulations
reported in \cite{Poppitz:2009gt} suggest the theory does maintain its consistency as a standard
quantum field theory would, and some newer result reported in \cite{Chen:2012di} 
is also of great interest.

\section*{Acknowledgements}
This research was supported by the John Templeton Foundation through
Professor John Moffat and in part by the Perimeter Institute for Theoretical 
Physics. Research at the Perimeter Institute is supported by the Government 
of Canada through Industry Canada and by the Province of Ontario through the 
Ministry of Economic Development and Innovation.

\appendix

\section{The non-abelian ``Stoke's theorem''}\label{app:stokes}
Let $\mathbb D\subset\mathbb L$ be a sub-lattice of $\mathbb L$ formed
by a certain set of plaquettes, and $\mathcal C=\partial \mathbb D$ be its boundary.
We prove Lemma \ref{lem:non-abelian_stokes} by 
induction on the number of plaquettes, $n$, contained in $\mathbb D$. When 
$n=0$, $w_{\mathcal C, x}=\mathbf 1$ and the Lemma obviously holds.
Suppose it is also true for $n=n_0$, we
consider the case for an arbitrary sub-lattice $\mathbb D$ that 
contains $n_0+1$ plaquettes.  

First, we note that the statement to prove is independent from
the choice of the base point. If $x$ and $x'$ are two
different vertices on $\mathcal C$ and $w_x$ and $w_{x'}$
are the Wilson-lines along $\mathcal C$ based at them
respectively. We know that $w_{x}=s\, w_{x'}\, s^{-1}$, where
$s\in G$ is the Wilson-line along the section of $\mathcal C$
connecting $x'$ to $x$. If the Lemma holds true for $w_{x'}$ and
so $w_{x'}=\mathcal P\prod_{p\in\mathbb D} c'_p f(p) c^{\prime -1}_p$,
by defining $c_p=s c'_p$, we find
$w=\mathcal P\prod_{p\in\mathbb D} c_p f(p) c^{-1}_p$.

Therefore, for the purpose of proving the Lemma, we may
choose an arbitrary base point of convenience. Given $n_0+1\ge 1$, it
is always possible to find a point $x$ on $\mathcal C$, such that
the immediate next link in $\mathcal C$ starting from $x$ is $l_2(x)$ 
and the plaquette, denoted by $p_0$, bounded by the vertices $x$,
$x+\hat 1$, $x+\hat 1+\hat 2$ and $x+\hat 2$ is contained in $\mathbb D$. 
For brevity, let us denote the gauge fields on the links
from $x$ and surrounding $p_0$ in the counter-clockwise direction as
$u_1$, $u_2$, $u_3$ and $u_4$, and so $f(p_0)=u_1 u_2 u_3 u_4$.

Consider the sub-lattice $\mathbb D'$ that contains
$n_0$ plaquettes obtained by removing the plaquette $p_0$ from $\mathbb D$.
The Wilson-line $w'$ around its boundary based at $x+\hat 1$, by
the induction hypothesis, is given by
\[
w'=\mathcal P \prod_{p\in \mathbb D'} c'_p\, f(p)\, c^{\prime -1}_p\,,
\]
where $c'_p$ is the Wilson-line along some path that connects the base point
$x+\hat 1$ to that of the plaquette $p\in\mathbb D'$.

Independent from how
the loop $\mathcal C$ may close itself at the point $x$, given
the above assumptions, it is evident that
\[
w=     u_4 u_3 u_2\, w'\, u_1=f(p_0)\, u_1^{-1} \,w'\, u_1\,.
\]
Define $c_p=u_1^{-1} c'_p$ for $p\in\mathbb D'$ and $c_{p_0}=\mathbf 1$, and
they obviously are the Wilson-lines along some paths that connect the point
$x$ to the base point of each plaquette in $\mathbb D$. Surely
\[
w=\mathcal P\prod_{p\in\mathbb D} c_p f(p) c_p^{-1}\,,
\]
which completes the proof.

\end{document}